# Efficient Approximation of Action Potentials with High-Order Shape Preservation in Unsupervised Spike Sorting

Majid Zamani, Christian Okreghe, and Andreas Demosthenous

*Abstract*— This paper presents a novel approximation unit added to the conventional spike processing chain which provides an appreciable reduction of complexity of the high-hardware cost feature extractors. The use of the Taylor polynomial is proposed and modelled employing its cascaded derivatives to non-uniformly capture the essential samples in each spike for reliable feature extraction and sorting. Inclusion of the approximation unit can provide 3X compression (i.e. from 66 to 22 samples) to the spike waveforms while preserving their shapes. Detailed spike waveform sequences based on in-vivo measurements have been generated using a customized neural simulator for performance assessment of the approximation unit tested on six published feature extractors. For noise levels $\sigma_N$ between 0.05 and 0.3 and groups of 3 spikes in each channel, all the feature extractors provide almost same sorting performance before and after approximation. The overall implementation cost when including the approximation unit and feature extraction shows a large reduction (i.e. up to 8.7X) in the hardware costly and more accurate feature extractors, offering a substantial improvement in feature extraction design.

## I. Introduction

Advances in microtechnology have enabled precise neural interaction monitoring of action potentials or spikes. The information carried by spikes has led to the development of a miniaturized and implantable intermediate deciphering processing unit called 'spike sorting' which separates individual spikes in clusters. Sorting of the spike waveforms helps to identify the optimal patterns and parameters to condition diseases by artificially modulating irregular or faulty electrical impulses [1], realizing a communication bridge for control of assistive devices for patients with damaged sensory/motor functions (e.g. hand prosthesis [2]) and to stimulate a particular pathway for biological functionality regularization [3].

There is a recent trend to develop high-channel count neural interfaces that include tens of thousands of sensing probes [4]. Large-scale data streaming to a remote computer for processing is challenging due to the data size (i.e. Gbps) and limits the viability of experiment set-ups. Compression techniques such as compressed sensing [5] and deep autoencoder [6] can be used for significant compression for data telemetry. However, they have the disadvantage that the data has to be recovered for further processing. The optimal solution is to perform ultra-low cost and highly efficient on-chip spike sorting [7]-[11]. As shown in Fig. 1(a), a conventional spike sorting chain comprises the following steps: 1) *detection and alignment*, separating spikes from noise and aligning the spikes to a common point, 2) *feature*

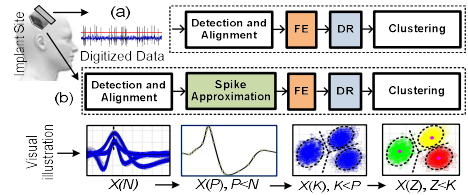

Fig. 1. (a) Conventional spike sorting for determining single unit activity. (b) Spike sorting with spike approximation. This processing chain performs almost loss-less approximation preserving the waveform shape before feature extraction (FE) and dimensionality reduction (DR). The overall sorting cost is reduced based on the approximation ratio.

*extraction (FE)*, extracting features of the spike shapes which gives a *dimensionality reduction (DR)*, i.e., reducing a space of dimension $N$ (the number of datapoints per spike) to a low dimensional space of a few features ($K$), and 3) *clustering*, grouping spikes with similar features into clusters ($Z$), corresponding to the different neurons. In Fig. 1(a) compression is proportional to ($N > K > Z$).

This paper proposes the inclusion of a high-order spike approximation unit before FE as shown in Fig. 1(b). It provides almost loss-less compression $P$ (e.g. $P = 1/3$) while preserving the geometrical shape of the spike waveforms (i.e. ($N > P > K > Z$)). Spike waveforms are applied to the approximation unit and only samples necessary to preserve the shape characteristic are retained, reducing the overall hardware implementation cost of the subsequent sophisticated and highly complex FEs. A Taylor polynomial [12]-[13] is used for spike approximation because of its strong representation capability and generalisation to the wide variety of spike shapes. It is modelled and implemented using consecutive time derivatives of aligned spike waveforms. It is demonstrated that: 1) Only three derivatives of the spike waveforms are needed to provide useful compression without compromising the spike sorting accuracy. 2) A Taylor polynomial features an asymmetric model structure for spike approximation, which is capable of leveraging low to high order geometrical information of spikes. 3) Introducing the approximation unit results in high-cost optimization of the spike sorting hardware, especially in the multiplication terms.

## II. High Order Taylor Polynomial and Spike Approximation

### A. Approximation Using Cascaded Derivatives:

Spike waveforms represent complex functions by concatenation of several morphological changes including

M. Zamani, C. Okreghe and A. Demosthenous are with the Department of Electronic and Electrical Engineering, University College London, Torrington Place, London WC1E 7JE, UK. (e-mail: m.zamani@ucl.ac.uk, christian.okreghe.16@ucl.ac.uk, a.demosthenous@ucl.ac.uk).

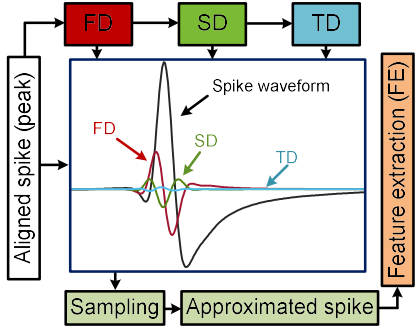

Fig. 2. Illustration of the proposed approximation method. A peak aligned spike is sent to the cascaded derivatives including FD, SD and TD. A sampling unit uses the derivatives information to approximate the spike waveform to a lower dimension by identifying the most relevant samples. The approximated spike is then used for FE.

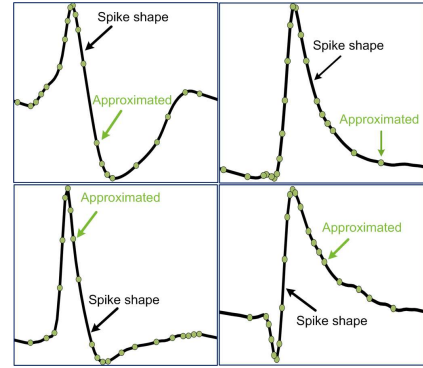

Fig. 3. Compression of the typical spike waveforms and their approximated versions. Four spike templates are used to show the effectiveness of the proposed approximation method. Samples are non-uniformly distributed along each spike to preserve the shape with 3X fewer samples, from $N=66$ to $P=22$.

slopes, curvature and amplitude variations, all of which are dependent on the observation point relative to the neuron. Therefore, information related to the morphological construction of spikes is essential to achieve good approximation and efficient compression at the same time. The Taylor polynomial [12]-[13] performs a high-order and robust characterization of spike waveforms by mapping them to a series of low to high order derivatives:

$$f(x) = f(a) + f'(a)(x-a) + \frac{1}{2!}f''(a)(x-a)^2 + \frac{1}{3!}f'''(a)(x-a)^3 + \cdots \quad (1)$$

where $f(x)$ is a function of a continuous variable that is differentiable up to order $\kappa + 1$ derivatives, and the expansion in (1) shows the approximation of $f(x)$ near a. The derivative terms in (1) are the indicators of the segments that can be exploited for non-uniform approximation (or sampling) of the initial function $f(x)$. The first derivative ($FD = f'$) of any function is described by its rate of slope variations and is computed as the difference between the current and previous sample point $FD = s(n) - s(n-1)$, where $s$ is the spike waveform and n is the sample number. $FD$ is shown in red in Fig. 2 and it highlights the gradient variations of a spike for retention of a subset of samples. As shown in Fig. 2, the output of the $FD$ is used to derive the second derivative $f'' = SD = FD(n) - FD(n-1)$ which is essentially associated with rate of change of the slope, representing the curvature of the signal. The third derivative is $f''' = TD = SD(n) - SD(n-1)$, which is effectively correlated to the higher order components that cannot be approximated by convex approximations. As the Taylor polynomial takes non-polynomial functions and maps them by using polynomial terms, the derivatives can be cascaded to any arbitrary order for higher degree approximation. This paper approximates the spike waveform using three derivatives only as shown in Fig.2.

*B. Spike Waveform Sampling*:

The absolute value of the spike derivatives together with their identified multiple peaks are considered for sampling. For example, the peaks of the first derivative listed in an ascending order are $(FD_{max1}, FD_{max2}, FD_{max3}, \cdots)$, $(SD_{max1}, SD_{max2}, SD_{max3}, \cdots)$ represent the recognized peaks of the second derivative and eventually $(TD_{max1}, TD_{max2}, TD_{max3}, \cdots)$ are the third derivative identified peaks. They clearly highlight the segments that can be used in re-sampling process and establish different sampling combinations. However, three sampling criteria should be taken into consideration to maximize the approximation efficacy:

- The curvature of the spikes should be preserved. Typically, the differences between high or low amplitude, and negative or positive curvatures make the spike waveform distinct. Therefore, the $(SD_{max1}, \cdots)$ should be used to fully capture the curvatures.
- The sampled curvatures should be connected using intermediate samples. $(FD_{max1}, \cdots)$ are the indicators of these samples.
- Accurate sampling of curvatures consisting of higher order convex components using $(TD_{max1}, \cdots)$ are the indicators of these samples.

Out of the wide range of combinations examined, the combination which strikes a good trade-off between spike approximation and sorting performance, and exhibits good noise immunity has three samples around $(SD_{max1}, SD_{max2}, SD_{max3}$ and $SD_{max4})$, one sample at $(FD_{max}, FD_{max2}, \cdots FD_{max6}$ and $FD_{max})$ and one sample at $(TD_{max1}, TD_{max2},$ and $TD_{max3})$. Overall, 22 out of 66 initial samples are saved which provides 3X compression of the original waveform while preserving the structure of the spike shape which is important for extraction of informative features in the FE unit. Examples of typical spike shapes and their approximations are shown in Fig. 3.

*C. Feature Extractor (FE) Methods for Comparison*:

Six published FE methods have been used for accuracy-cost analysis when adding the approximation unit. There are two categories: simple on-chip FE methods [8]-[9], and complex off-line methods [14]-[16]. The latter help proper evaluation of the approximation unit in implementation cost optimization and comparison to the on-chip FE methods.

*C.1) Adaptive Discrete Derivatives (ADDs) [8]*:
ADDs are computed by calculating the slope at each sample point over a number of different time scales:

$$\text{ADDs} = amp[s(n) - s(n-\delta)|_{\delta=1\ldots7}] \quad (2)$$

where $amp = 1$ is the amplitude of the decomposition window, $s$ is the spike waveform, $n$ is the sample point, and $\delta$ is the scaling factor (time delay). Adjustment of the scaling factors ($\delta 1, \delta 2, \delta 3$) are based on three frequency sub-bands from $\delta = 1$ to $\delta = 7$ corresponding to the most deviated features (non-Gaussian features) for unsupervised clustering.

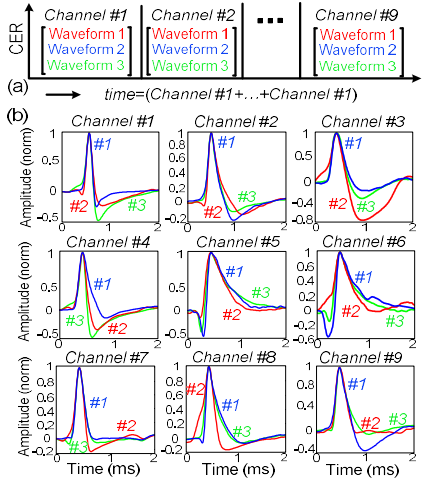

Fig. 4. (a) The spike waveforms in the recording channels (#1…#9) are changed over time to embed variety in the data stream. This ensures proper assessment of the approximation method under diverse conditions. (b) Corresponding spike mean waveforms in each recording channel with normalized amplitudes and 2 ms duration.

*C.2) Zero Crossing Features (ZCFs)* [9]: ZCFs are the integral of positive and negative lobes which normally contain the information about the amplitude variations of the spike lobes, position of the positive and negative spike peaks, and width of the lobes.

*C.3) Spike Shape:* In this method all the samples of detected and peak-aligned spikes (without upsampling) are used for calculating the similarity measure between the mean of the spikes in clustering.

*C.4) Discrete Wavelet Transform (DWT)* [14]:
Four-level multi-resolution decomposition using Haar wavelets is calculated as feature extraction which results in 64 wavelet coefficients for each spike. Then, 10 features with the most deviation from normality are extracted for the sorting stage.

*C.5) Updated Graph Laplacian Features (uGLF)* [15]: GLF is a linear feature extraction technique that simultaneously minimises the graph Laplacian and maximises variance. The important attribute of GLF is that the points which are close to each other in high dimensional space remain close to each other after transformation to low dimensional space to ensure the clusters are compact and separable. In GLF, a weighted graph representing the projection matrix is constructed and updated according to the recording channel spike waveforms in the neural simulator.

*C.6) Updated Principal Component Analysis (uPCA)* [16]: In uPCA, the feature extraction matrix changes over time, adapting itself to the recorded data information and projecting the spike waveforms in a lower dimensional space while preserving the main characteristics.

*D. Neural Data Simulator*:

To generate neural spikes, a database of synthetic spike waveforms containing 300 different average spike shapes was constructed. The spike shapes were extracted from the peripheral median nerve in pig (obtained with a multi-electrode cuff in vivo) [10], the neocortex and basal ganglia, from the right and left hippocampus and either from the right or left amygdala. Fig. 4 shows the procedure used for neural data generation and examination of the proposed approximation method and the FE algorithms. The neural spike waveforms are selected and placed in a data stream from the spike library to emulate a real recording channel. As shown in Fig.4(a), there are nine channels interleaved (*Channel#1…Channel#9*) that each has three distinct spike waveforms (Waveform1…Waveform3). Each channel accommodates diverse spike waveforms which is extremely important in approximation robustness evaluation before FE and sorting. They are also corrupted by additive noise with varying standard deviations (e.g. $\sigma_N = 0.01$) to examine the ability of the approximation unit to identify the informative spike samples. Classification error is defined as $(CER = (1 - CA_{CC}))$, where $CA_{CC}$ is the number of truly assigned feature vectors over the total number of feature vectors. CER is calculated over time considering the number of interleaved channels $time = (Channel\#1 + \cdots + Channel\#9)$.

## III. RESULTS AND DISCUSSION

The K-means function in Matlab with the number of iterations set to 10 for near-optimum CER analysis was used. Fig. 5(a) shows the CER variations of the spike shapes (*N*=66) and Fig. 5(b) their approximated version (*N*=22) using the neural simulator outputs. Varying the spike waveform shapes in the interleaved recording channels ($Channel\#1 + \cdots + Channel\#9$) and noise levels $\sigma_N = (0.05, 0.1, 0.15 \text{ and } 0.2)$ has negligible effects on the CER (almost zero%). The template and noise robustness confirm that the proposed approximation method successfully removes the unnecessary samples in each spike. Fig. 6 shows the CER versus the computational complexity of the FEs. The computational complexity [8], [10] is defined as $(Comp = N_{add/sub} + 10 \times N_{mul/div})$, where $N_{add/sub}$ is the number of additions (or subtractions), and $N_{mul/div}$ is the number of multiplications (or divisions) required. There is a reduction in complexity shown in Fig. 6 in DWT, uGLF and uPCA on application of the approximation unit. This is due to optimizing the complexity of multiplications (or divisions) that are 10 times more costly than additions (or subtractions) when using the proposed approximation unit. The multiplication cost of uPCA is $kW(N^2 + N)$ where $k = 9$ is the number of channels, $W = 3$ is the number of waveforms in each channel and $N$ is the number of samples per spike. The approximation unit simplifies and reduces the FE cost by selecting the appropriate samples, in this work from 66 to 22 which means 3X compression. Therefore, the uPCA implementation cost is diminished by 8.7X from $uPCA_{66} = 1193940$ to $uPCA_{22} = 136620$ which is of the same order as the 10X multiplication cost. It should be noted that the cost of on-chip FEs (ADDs, ZCFs and spike shape) are not affected by the approximation method, because the reduced FE cost is compensated by the calculation of the derivatives. Overall, inclusion of the approximation method which changes the more accurate complex FEs to their much lower cost versions and may enable them to be included in on-chip processing. Further reduction of the spike samples using other approximation methods will be explored in the next studies.

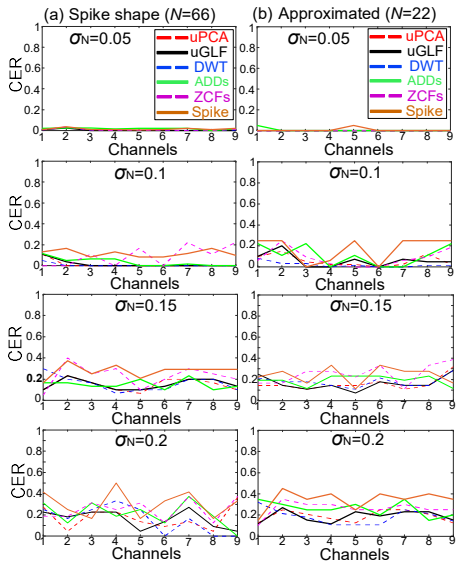

Fig.5. Comparison of classification error (CER) of feature extractors versus noise level $\sigma_N = (0.05, \ldots, 0.2)$ for the waveforms across the channels ($Channel\#1 + \cdots + Channel\#9$). (a) $N=66$ samples and (b) 22 samples after approximation. CER is almost preserved by varying the recording channels information over time.

## IV. CONCLUSION

In this paper, a new approximation method in spike sorting based on the Taylor polynomial has been proposed. It uses the first three derivatives to identify the eligible sampling segments. Selection criteria were also formulated for non-uniform approximation of the spike waveforms. Approximation of the spikes specifically selects the curvatures, intermediate connecting samples and the higher order convex components. For detailed performance assessment of the approximation method a customized neural simulator was designed consisting of nine channels to emulate the neural recording where each channel has unique spike waveforms. The CER of different FEs were tested using the K-means classifier and it was demonstrated that the proposed approximation unit retains and preserves the critical samples in a spike waveform without compromising the CER. The 3X sample reduction results in complexity reduction for DWT, uGLF and uPCA. For example, the CER of DWT before and after spike approximation is 0.01, and the 3X approximation reduces the DWT multiplication term $(8N - 10)$ by 60%. The added approximation unit in the sorting chain introduces a powerful tool for on-chip implementation of high-cost FE methods.

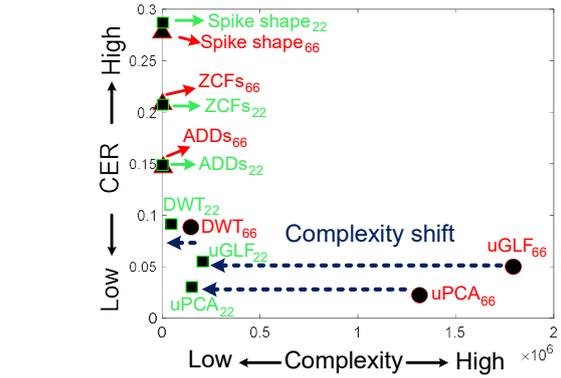

Fig. 6. Classification error (CER) versus computational complexity for the different FE methods. The use of the approximation unit does not introduce additional cost to the on-chip FEs (ADDs, ZCFs and spike shape) as there is almost zero shift before and after approximation (i.e. ADDs$_{66}$ → ADDs$_{22}$). However, there is major complexity reduction for the complex FEs after approximation (DWT$_{66}$, uGLF$_{66}$ and uPCA$_{66}$) to (DWT$_{22}$, uGLF$_{22}$ and uPCA$_{22}$). CER is almost the same due to the shape preservation of the spikes.